\documentclass[12pt]{article}
\usepackage{osajnl,overcite}				
\usepackage{amsmath,amsfonts,amssymb,graphicx}
\usepackage{mathrsfs,bbm}				
\usepackage{pst-all,pstricks-add}			

	\newcommand{\E}[1]{\ensuremath{\mathbb{E}\!\left[#1\right]}}

	\newcommand{\norm}[1]{\ensuremath{\|#1\|}}

	\newcommand{\abs}[1]{\ensuremath{\left|#1\right|}}

	\DeclareMathOperator{\Var}{Var}

	\DeclareMathOperator{\cov}{Cov}

	\newcommand{\mean}[1]{\langle #1 \rangle}

	\newcommand{\brho}{\boldsymbol{\rho}}
	
	\newcommand{\Bs}[1]{\boldsymbol{#1}}
	
	\renewcommand{\Re}{\ensuremath{\mathbb{R}}}

	\DeclareMathOperator{\Sch}{\mathcal{S}(\Re)}
	\DeclareMathOperator{\Scd}{\mathcal{S}^\ast(\Re)}

	\newcommand{\etal}{\textit{et al.}}
	
	\DeclareMathAlphabet{\mathpzc}{OT1}{pzc}{m}{it}

\begin{document}

\title{Modeling turbulent wave-front phase as a fractional Brownian motion: a new approach.}

\author{Dar\'{\i}o G. P\'erez, Luciano Zunino and Mario Garavaglia}
\affiliation{Centro de Investigaciones \'Opticas (CIOp), \\ CC. 124 Correo Central, La Plata 1900, Argentina.}

\begin{abstract}
This paper introduces a general and new formalism to model the turbulent wave-front phase using fractional Brownian motion processes. Moreover, it extends results to non-Kolmogorov turbulence. In particular, generalized expressions for the Strehl ratio and the angle-of-arrival variance are obtained. These are dependent on the dynamic state of the turbulence.
\end{abstract}

\section{INTRODUCTION}\label{sec:intro}

Earth turbulent atmosphere introduces spatial and temporal variations in wave-front that lead to image degradation of optical systems. Astronomical telescopes, laser beam projection systems, and optical communication systems are limited by the presence of turbulence. In particular, the resolution of a ground-based telescope is notably modified. Generally, the telescope aperture is assumed to be smaller than the outer scale of the turbulence, so spatial frequencies of the turbulence with wavelength of the order of the aperture diameter, $D$, impart a random tilt on the incident wave-front. This wave-front tilt translates to simple image motion at the image plane. It is the dominant atmospheric aberration across the telescope pupil. Statistical characterization of the image motion is of paramount importance because of its implications on the design of adaptive optics systems. 

In order to characterize temporally and spatially the statistics of the wave-front phase $\varphi$ several sensing methods have been used \cite{paper:silbaugh}. They use single\cite{paper:acton}, double\cite{paper:fried,paper:sarazin}, and multiple\cite{paper:dayton,paper:nicholls} (Shack-Hartmann) aperture sensors to measure the wave-front tilt.  The centroid of the long-exposure images formed by each aperture is directly proportional to the slope of the wave-front across it.

As usual, the phase structure function is
\begin{equation}
D_{\varphi}(\brho'-\brho)=\mean{\abs{\varphi(\brho')-\varphi(\brho)}^2},
\label{def:struct}\end{equation}
where $\brho',\brho\in\Re^2$, and $\mean{\cdot}$ stands for the average using some unknown probability distribution\cite{note:structure-function}. 
Whenever a Kolmogorov developed turbulence is present, under the small perturbation and near-field approximations, the latter   is turned into the widely known\cite{paper:rodier}
\begin{equation}
D_{\varphi}(\brho'-\brho)= C^2_\varphi \left(\frac{\norm{\brho'-\brho}}{r_0}\right)^{5/3},
\label{eq:original}
\end{equation}
where $r_0$ is the Fried parameter\cite{paper:fried-2} linked to the spatial statistical properties of the refractive index, and $C^2_\varphi$ is the phase structure constant, roughly near to  $6.88$. 

Interferometric measurements have corroborated the expression in Eq.~\eqref{eq:original}. Many of these measurements have been made under the conditions mentioned above. But, significant departures from the $5/3$ exponent have been experimentaly observed\cite{paper:acton,paper:dayton,paper:nicholls,paper:buser,paper:bester}. 

In particular, for near to the ground measurements, exponents in the range $(1,5/3]$ have been determined experimentally. It is well-known that atmospheric turbulence is not always in its fully developed state, thus deviations from this simple model are likely---non-Kolmogorov turbulence. The phase structure function can then be generalized\cite{paper:nicholls,thesis:stribling} to include these results as follows,
\begin{equation}
D_{\varphi}(\brho'-\brho)=C^2_{\varphi,\beta} \left(\frac{\norm{\brho'-\brho}}{r_{0,\beta}}\right)^{\beta - 2},
\label{eq:modified}\end{equation}
where $\beta$ is the exponent associated with the phase spectrum, $r_{0,\beta}$ is the generalized Fried parameter and $C^2_{\varphi,\beta}$ is a constant maintaining consistency between the power spectrum and the structure function of phase fluctuations. If a Kolmogorov spectrum is chosen: $\beta=11/3$, $C^2_{\varphi,\beta} \approx 6.88$ and   $r_{0,\beta}=r_0$; thus, we recover Eq.~\eqref{eq:original}.

In order to model turbulence-degraded wave-fronts by Kolmogorov turbulence, Schwartz \etal\, \cite{paper:schwartz-baum} have suggested that these are fractal surfaces described by a fractional Brownian motion (fBm) process with Hurst parameter $5/6$ and a fractal dimension equal to $13/6$. Fractal properties are attributed to both the spatial and temporal behavior, and they are directly related through the Taylor hypothesis, or frozen turbulence approximation. The value of the Hurst parameter is in accord with the predictability of real stellar wave-front slopes\cite{paper:jorgenson-2}. Moreover, several algorithms for adaptive optics have been designed based on this statistical prediction \cite{paper:schwartz-baum,paper:jorgenson-2,paper:lloyd-hart}. In particular, McGaughey \& Aitken \cite{paper:mcgaughey} have suggested that the source of predictability stems from the low-pass spatial filtering done by the wavefront sensor. These authors have modeled the temporal behavior of wavefront slopes by filtering the increments of a fBm with Hurst parameter equal to $5/6$.

On the other hand, several authors have simulated Kolmogorov phase fluctuations over a finite aperture by using the random midpoint displacement algorithm\cite{paper:lane,paper:lakhal,note:aitken}. This is also an algorithm for  generating fractional Brownian motion surfaces---see Ref.~\cite[pp. 487--490 and 497--499]{book:peitgen}. Moreover, the fBm model was used as an alternative stochastic process to model the turbulent refractive index in lightwave propagation \cite{paper:perezzunino}. 

In this paper the turbulent wave-front phase is modeled as a fBm. The result of Schwartz \etal\ is extended to include non-Kolmogorov turbulence. In Sec. \ref{sec:fbm} the general formalism for fBm processes is presented. This formalism is applied to determine the Strehl ratio in Sec. \ref{sec:app}.\ref{sec:strehl} and the angle-of-arrival variance in Sec. \ref{sec:app}.\ref{sec:variance}. The latter result is compared against the well-known and widely used variance first obtained by Tatarsk\u{\i}.

\section{FRACTIONAL BROWNIAN MOTION AND ITS ASSOCIATED NOISE}\label{sec:fbm}

Usually, natural phenomena behaving randomly are labeled as \textit{noises}. These noises are characterized through the estimation of their power spectrum, $\mathcal{W}(\nu)$\cite{paper:chen,book:wornell,book:peters}. Empirically, an enourmously wide range of these spectra have been observed to follow power-laws proportional to $\abs{\nu}^{-\beta}$, for some exponent $\beta$. Better known as $1/f^\beta$-type noises\cite{note:frequency}; they are classified according to the value of the exponent, e.~g. Ref.~\cite[ch. 3]{book:wornell}. 

Since its first formalizations in the earlies 1900's (independently modeled by L. Bachelier and A. Einstein) the Brownian motion caught the attention of physicists. It is the most common representative of $1/f^2$-type noises; thenceforth, processes with such spectra are known as \textit{brown noises}. On the other hand, the derivative of the Brownian motion is called \textit{white noise}. The fact that it only can been defined as a distribution (in some probability space) is found in its tail-divergent power spectral distribution, i.~e. $\beta=0$. Afterwards, any process between these two, with power exponent $0 < \beta <2$, is referred as a \textit{pink noise}. The last category is for those processes with $2<\beta<3$, they are considered \textit{black noises.} 

A (stochastic) process $X(t)$ is \textit{self-similar} with index $H$ if, for any $c>0$,
\begin{equation}
X(t)\overset{d}{=}c^H X(c^{-1}t);
\label{eq:H-ss}\end{equation}
that is, both processes are equal in distribution. The coloured noises are self-similar, with exponent $H = (\beta -1)/2$, as the generalized Fourier transform of their spectra can show\cite{lecture:flandrin}. Also, it suggests the presence of a slowing decaying auto-correlation, and thus of memory.

Nevertheless, this colour classification is rather rough. It is insufficient knowledge of the power spectra to create stochastic processes modeling the randomness of quantities observed in the real world. As these random quantities tend to appear in dynamics equations other properties are needed to give them sense, e.~g.: bimanual rhythmic coordination differential equation\cite{paper:chen}, Black-Scholes market equation\cite{paper:elliot}, ray-optics equation\cite{book:th-dario}, etc..

Because of all the `good' properties it endows, stationarity is desired. It is said a processes $X$ is \textit{(wide sense) stationary} if
\begin{align*}
\mean{X(t)}=\mean{X(0)} &\text{ and } \cov (X(t+\tau),X(t))=\cov (X(\tau),X(0)),
\end{align*}
for any $\tau\in\Re$. 

Unfortunately, self-similar processes are non-stationary: $\Var X(t)\propto t^{2H} \neq$ const.. Since we have lost ergodicity, the \textit{Wiener-Khinchin theorem} fails. On the other hand, the existence of stationary increments does not contradicts the self-similar property. A process with \textit{stationary increments} is such that the probability law of its increments $X(t+\tau)-X(t)$ is independent of $t$. 

Natural phenomena exhibit in general a non-gaussian behavior. Nevertheless, it is usual to append these to a gaussian distribution since, in this way, they become analytically tractable, Ref.~\cite[p. 35]{book:wornell}. Moreover, choosing this distribution leaves unaffected the memory properties described by the spectrum. That path will be followed here. 

There is only one family of processes which are self-similar, with stationary increments, and gaussian: the\textit{ fractional Brownian motion} (fBm)\cite{book:taqqu}. The normalized family of these gaussian processes, $B^H$, is the one with\cite{paper:mandelbrot-2}
\begin{equation*}
B^H(0)=0\quad\text{ almost surely, }\quad\mathbb{E}[B^H(t)]=0,
\end{equation*}
and  covariance
\begin{equation}
\E{B^H(t)B^H(s)}=\frac{1}{2}\left(\abs{t}^{2H}+\abs{s}^{2H}-\abs{t-s}^{2H}\right),
\label{eq:fbm-cov}\end{equation}
for $s,t\in\Re$. Here $\E{\,\cdot}$ refers to the average with gaussian probability density. The power exponent $H$ is the \textit{Hurst parameter} and its range is bounded. While the condition $H>0$ guarantees their (mean-square) continuity, $H<1$ avoids degeneracy\cite{lecture:flandrin}. Another more intuitive argument can be drawn. It is well-known these curves have fractal dimension equal to $2-H$\cite{book:falconer}. Because they are embedded in the plane, $H>0$. On the other hand,  continuous parameterized curves should have dimension greater than one, and thus $H<1$.

These processes exhibit memory, as can be observed from Eq.~\eqref{eq:fbm-cov}, for any Hurst paremeter but $H=1/2$. In this case successive Brownian motion increments are as likely to have the same sign as the opposite, and thus there is no correlation. Otherwise, it is the Brownian motion that splits the family of fBm processes in two. When $H>1/2$ the correlations of successive increments decay hyperbolically, and this sub-family of processes have long-memory. Besides, consecutive increments tend to have the same sign, these processes are \textit{persistent}. For $H<1/2$, the correlations of the increments also decay but exponentially, and this sub-family presents short-memory. But since consecutive increments are more likely to have opposite signs, it is said that these are \textit{anti-persitent}.

Fractional Brownian motions are continuous but non-differentiable processes (in the usual sense), and only give spectra exponents between $1$ and $3$. Nevertheless, fBm processes can be generalized to allow derivatives. A simple dimensional inspection suggests that the latter should have spectral exponent equal to $\beta=2H-1$; thus, covering the range $-1<\beta<1$. 

Formally, continous processes are not called noises since they can be integrated \textit{pathwise}. That is, given any continuous process $X(t)$ the integral
\begin{equation*}
\int^b_a X(t) B^H(t)\, dt
\end{equation*}
exists for any realization of the integrands. On the other hand, noises are not pathwise integrable. That is, the integral
\begin{equation*}
\int^b_a X(t) \frac{d}{dt} B^H(t)\, dt := \int^b_a X(t)\, d B^H(t)
\end{equation*}
is not the limit of area approximating sums for any realization, i.~e., there is no calculus in the classical sense.

The first construction of a \textit{Stochastic Calculus} was made by It\^o around 1940 for Brownian motions. Later, these results were extended to more general processes---semi-martingales\cite{book:oksendal} and infinite dimensional Wiener (Brownian) processes\cite{book:nualart}. The \textit{White Noise Analysis} due to Hida, focused in the white noise rather than the Brownian motion as a fundamental entity, is of particular interest here. 

Like common sense suggests the lack of conventional derivatives should be overcome through distributions. This is the basic idea underlying the white noise calculus. The problem is thus to embed these distributions into the right probability space. Let $\phi$ be an element of the \textit{Schwartz space} $\Sch$ (the space of rapidly  decreasing smooth real-valued functions), and $\omega$  is any element of the dual $\Scd$. Therefore, the \textit{white noise} is defined as the bilinear map $W$ such that, $W(\phi)= \langle\phi,\omega\rangle= (\phi,W(\omega))$---where $\langle\cdot,\cdot\rangle$ is the bilinear map and $(\cdot,\cdot)$ the usual internal product in $L^2(\Re)$. The space $\Scd$ turns out to be a gaussian probability space and its elements $\omega$ the events. Moreover, the pairing coincides with the It\^o integral, i.~e. $\langle\phi,\omega\rangle=\int_\Re \phi\, dB$\cite{book:holden}, and using its properties (Ref.~\cite[p. 15]{book:holden}) it is found:
\begin{equation*}
 (\phi,W(\omega))=\int_\Re \phi\, dB = -\int_\Re \frac{d\phi}{dt} B\, dt = (- d\phi/dt, B);
\end{equation*}
that is, the white noise, as it was defined, is the derivative of the Brownian motion. 

In the last decade different approaches have been given to extend the stochastic calculus to fBm. The range of persistent processes have been particularly fruitful\cite{paper:zaehle-1,paper:zaehle-2, paper:decreusefond,paper:follmer}, not only because of its applications in practical problems also for its regularity properties. Duncan \etal\,\cite{paper:duncan} successfully extended the white noise calculus to this range by means of a tool termed Wick product. These ideas were recently picked up by Elliott and van der Hoek\cite{paper:elliot} who have given a complete calculus for all values of the Hurst parameter. A brief outlook based on their work is given in Appendix A.

\section{WAVE-FRONT MODELING AND APPLICATIONS}\label{sec:app}

Let $\varphi$ be the phase difference between the average and perturbed wave-front. As it was argued at the introduction, it is a realization of a fractal surface. Moreover, the small perturbation and near-field approximations guarantee structures functions like Eqs.~\eqref{eq:original} or \eqref{eq:modified}. That is, it has stationary increments. Now, as always, it is assumed that the process $\varphi$ is gaussian---see for example Ref.~\cite[p. 293]{paper:rodier}. Its power spectrum is also observed to follow a power law; thus, it is self-similar. At least it is valid within the inertial range, which is limited by two characteristic scales---the outer and inner scales, $L_0$ and $l_0$ respectively. 

Now, let 
\begin{equation}
\tilde{B}^H(\brho) := B^H(\norm{\brho}) = B^H(\rho)
\label{eq:ifbm}\end{equation}
be the \textit{isotropic fractional Brownian motion} (ifBm). It is gaussian, self-similar and, under condition \eqref{eq:less-1/2} (given at the Appendix B), has stationary increments. Therefore, we can define the \textit{generalized} phase difference as
\begin{equation}
\varphi := C_\varphi \tilde{B}^H\!\left(\frac{\brho}{r_0}\right),
\label{eq:fbm-phase}\end{equation}
where $C_\varphi$ is defined as in Eq.\eqref{eq:original}, and $H=5/6$ in the Kolmogorov turbulence case. Its structure function is, 
\begin{align}
\E{\abs{\varphi(\brho')-\varphi(\brho)}^2} &=  C^2_{\varphi} \left(\frac{\abs{\rho-\rho'}}{r_0}\right)^{2H}\nonumber\\
&\simeq C^2_{\varphi} \left(\frac{\norm{\brho-\brho'}}{r_0}\right)^{2H},
\label{eq:struct-H}\end{align}
where the last step is made under the condition $\abs{(\rho-\rho')/r_0}^{3/2}\ll 1$, which guarantees this process has stationary increments---see Appendix B. Observe that since the Fried parameter can be interpreted as the diameter of the coherence area of the perturbed wave-front the last restriction is compatible. As it was stated earlier at the introduction, the structure function power exponent is restricted to the range $(1,5/3]$ for near to the ground measurements. Therefore, the Hurst exponent is confined to $1/2< H\leq 5/6$. 

\subsection{Strehl ratio}\label{sec:strehl}

The Strehl ratio, $S$, is used as a figure of merit in order to characterize the quality of beam propagation and the performance of a telescope. It is defined as the ratio of the central irradiance of the aberrated point spread function and the diffraction-limited point spread function\cite{note:tilt}.

Considering a circular aperture of diameter $D$ receiving an optical signal it was shown\cite{thesis:stribling}: 
\begin{equation*}
S=\frac{16}{\pi}\int_0^1 u \,du  \left[(\cos{u})^{-1}-u\sqrt{1-u^2}\right]\,\exp\left[{-\frac{1}{2}D_w(D\,u)}\right],
\end{equation*}
where $D_w(\rho)$ is the wave structure function. In the near-field approximation the wave structure function  is replaced by the phase structure function. By using Eq.~\eqref{eq:struct-H} leads to
\begin{equation}
S=\frac{16}{\pi}\int_0^1 u\,du\left[(\cos{u})^{-1}-u\sqrt{1-u^2}\right]\,\exp\left[{-\frac{C^2_{\varphi}}{2}\left(\frac{D}{r_0}\right)^{2H} u^{2H}}\right].
\end{equation}
In the case of Kolmogorov turbulence ($H=5/6$) the following well-known expression is recovered\cite{paper:roddier-2}:
\begin{equation*}
S=\frac{16}{\pi}\int_0^1 u\,du\left[(\cos{u})^{-1}-u\sqrt{1-u^2}\right]\exp\left[{-3.44\left(\frac{D}{r_0}\right)^{5/3}u^{5/3}}\right].
\end{equation*}
Remember that, for small phase aberration $\sigma_\varphi^2<<1$, the Strehl ratio can be expressed as a function of the phase variance \cite{book: born}: $
S \simeq \exp({-\sigma_\varphi^2}).$ This formula implies that the normalized intensity is independent of the nature of the aberration and is smaller than the ideal unity value by an amount proportional to the phase variance. Under the definition in Eq.~\eqref{eq:fbm-phase}, it is 
\begin{equation}
S \simeq \exp\left({-\frac{C^2_{\varphi}}{r^{2H}_0}\,\rho^{2H}}\right).
\end{equation}
Therefore, not only through the Fried parameter the quality of beam propagation can be set but also the Hurst parameter is relevant.

\subsection{Angle-of-arrival variance}\label{sec:variance}

The path difference, or wavefront corrugation, of the wavefront surface from the average plane is simply
\begin{equation}
z(\brho) =\frac{\lambda}{2\pi}\, \varphi(\brho),
\label{eq:z-path}\end{equation}
 as usual, $\lambda$ is the wavelength. Light rays are normal to the wavefront surface within the framework of Geometric Optics. The angle-of-arrival, at each normal plane, is
\begin{equation}
\theta_i= -\frac{\lambda}{2\pi}\,\frac{\partial \varphi}{\partial x_i}, 
\end{equation}
$(i=1,2)$. Following classical arguments, Ref.~\cite[pp. 334-336]{paper:rodier}, the angle-of-arrival variance is given by
\begin{equation}
\sigma_{m}^2=\mean{\theta_1^2}+\mean{\theta_2^2}=\int_\Re\! d^2\nu \left[\mathcal{W}_{\theta_1}(\boldsymbol{\nu})+\mathcal{W}_{\theta_2}(\boldsymbol{\nu})\right]= \lambda^2\int_\Re\!\nu^2 d^2\nu\;  \mathcal{W}_\varphi(\boldsymbol{\nu}),
\end{equation}
where $\mathcal{W}_\varphi(\nu)$ is the power spectrum of $\varphi(\boldsymbol{\rho})$. It should be stressed that in this expression it is applied the \textit{Wiener-Khinchin theorem}; thence, the phase is modeled as a stationary random variable.

A divergent integral is obtained under the assumptions of Kolmogorov turbulence, small perturbation and near field approximation. In order to make it summable Roddier introduces a high and low frequency cut-offs, $D^{-1}$ and $L_0^{-1}$ respectively. It is a `more realistic expression' where the aperture diameter and the turbulence outer scale 
 are involved. That is,
\begin{equation*}
\sigma_{m}^2\propto \lambda^2 r_0^{-5/3} \int_{L_0^{-1}}^{D^{-1}}\! d\nu\, \nu^{-2/3}.
\end{equation*}
We can obtain the following result, integrating the above equation and considering $D\ll L_0$: 
\begin{equation*}
\sigma_{m}^2\propto \lambda^2 D^{-1/3} r_0^{-5/3}.
\end{equation*}
A more precise relation was given by Tatarsk\u{\i}\cite{book:tatarskii-2}:
\begin{equation}
\sigma_{m}^2\simeq (6.88/{2\pi^2})\lambda^2 D^{-1/3} r_0^{-5/3},
\end{equation}
where the proportionality coefficient is in radians squared units. However, coefficients ranging from $0.342$\cite{paper:brandt,paper:acton-2}, $0.358$\cite{paper:fried,paper:sarazin}, to $0.365$ \cite{paper:olivier} have been given. It should be noted that these coefficients were obtained by using just only phase differences, so it is necessary that the wave-front remains unchanged over the whole aperture. Then, the pupil size must be smaller than the inner scale of the atmospheric turbulence, i.~e. $D<l_0$.

Now, according to Eqs.~\eqref{eq:fbm-phase} and \eqref{eq:z-path}, we have that  
\begin{equation*}
z(\brho) =   C_z \,\tilde{B}^H\!\left(\frac{\brho}{r_0}\right),
\end{equation*}
where $C_z= \lambda C_\varphi/2\pi$. Therefore, the angle-of-arrival is
\begin{equation}
\theta^H_i(\brho)= -\frac{\partial z}{\partial x_i} = - C_z r_0^{-H} \frac{W^H(\rho)}{\rho} x_i,
\label{eq:AOA}\end{equation}
where $W^H$ is the fractional white noise as defined at Appendix A. The total variance of the angle-of-arrival is  
\begin{equation*}
\sigma_{m,H}^2(\rho)=\E{\theta^2_1 + \theta^2_2}(\rho) = C^2_z r_0^{-2H}  \E{(W^H(\rho))^2}.
\end{equation*}
Let us calculate the fractional white noise variance using its chaos expansion, Eq.~\eqref{eq:fwn-exp}, and the Wick product in Eq.~\eqref{eq:wick-gauss}. Since, it is
\begin{equation*}
W^H(\rho) W^H(\rho') =W^H(\rho) \diamond W^H(\rho') + \sum^{\infty}_{n=1} M_H\xi_n(\rho) M_H\xi_n(\rho'),
\end{equation*}
and considering statistical dependent variables are treated as if they were independent with respect to the average when Wick multiplied, $\E{W^H\diamond W^H}= \E{W^H}\cdot\E{W^H}= 0\cdot 0 = 0$---the last step is for the noise being a zero-mean gaussian variable. We have the following 
\begin{equation}
\E{W^H(\rho) W^H(\rho')} = \sum^{\infty}_{n=1} M_H\xi_n(\rho) M_H\xi_n(\rho').
\label{eq:wfbn-variance}\end{equation}
Finally, from Eq.~\eqref{eq:AOA} and the latter equation:
\begin{equation*}
\sigma_{m,H}^2(\rho) =  C^2_z r_0^{-2H} \sum^{\infty}_{n=1} M_H\xi_n(\rho) M_H\xi_n(\rho).
\end{equation*}
The sum in the above equation can be analytically calculated. Consider
\begin{equation*}
\widehat{M_H\xi_n}(\nu) = c_H \abs{\nu}^{1/2-H}\hat{\xi}_n(\nu)= - c_H\abs{\nu}^{1/2-H} i^{n-1}\xi_n(\nu),
\end{equation*}
where the Fourier transform\cite{note:ft} property of the Hermite functions was employed. Therefore, 
\begin{align}
\sum^{\infty}_{n=1} M_H\xi_n(\rho) M_H\xi_n(\rho') =& \frac{c^2_H}{2\pi}\int_{\Re^2}d\nu\,d\nu' \abs{\nu}^{1/2-H}\abs{\nu'}^{1/2-H} \left[\sum^{\infty}_{n=1} (-1)^{n-1}\xi_n(\nu) \xi_n(\nu')\right] \exp i(\nu \rho+\nu' \rho')\nonumber\\
=& \frac{c^2_H}{2\pi}\int_{\Re^2}d\nu\,d\nu' \abs{\nu}^{1/2-H}\abs{\nu'}^{1/2-H} \delta(\nu+\nu') \exp i(\nu \rho+\nu' \rho')\nonumber\\
= &\frac{c^2_H}{2\pi}\int_{\Re}d\nu\,\abs{\nu}^{1-2H} \exp i\nu (\rho- \rho'),
\label{eq:inf-sum}\end{align}
for the intermediate steps one has to use the orthogonality and parity of the Hermite functions. First note from this equation that the covariance is stationary. But, if we set $\rho=\rho'$ the angle-of-arrival variance
\begin{equation*}
\sigma_{m,H}^2= C^2_z r_0^{-2H} \frac{c^2_H}{2\pi}\int_\Re d\nu\,\abs{\nu}^{1-2H}
\end{equation*}
is divergent! Then let us follow Roddier's idea and introduce an adequate cut-off to the above,
\begin{align*}
\sigma_{m,H}^2&= \frac{\lambda^2}{(2\pi)^2} C^2_\varphi r_0^{-2H} \frac{ c^2_H }{\pi}\int^{2 D^{-1}}_0 \!\!\!d\nu\, \nu^{1-2H}\\
&=\frac{\Gamma(2H+1)\sin \pi H}{2^{2H}\pi(1-H)}\, \frac{C^2_\varphi}{2\pi^2}\lambda^2 r_0^{-2H} D^{2H-2}.
\end{align*}
Observe that for $H$ equal to $5/6$, it is $\sigma^2_{m,5/6}= 0.452552\, \sigma_m^2$---where $\sigma_m^2$ is the variance obtained by Tatarsk\u{\i}. The cut-off \textit{\`a la} Roddier notably reduces the value estimated by Tatarsk\u{\i} and others. As it was pointed out earlier, the scales considered throughout this paper are above the inner scale: in particular, $D>l_0$. Then, the difference between these variances is plausible. 

As we have seen, the removal of high frequencies is due to the finite size of the aperture\cite{paper:conan}. In fact, since many scales are involved this filtering must be introduced in order to smooth out. Let us properly introduce this effect. 

Define the \textit{smoothed fractional white noise} as follows: given $\phi_\rho(s)=\phi(s-\rho)$, \begin{equation}
W^H_\phi(\rho) = (\phi_\rho, W^H)=\sum^{\infty}_{n=1} (\phi_\rho, M_H\xi_n) \mathcal{H}_{\epsilon_n}(\omega)
\label{eq:smooth-noise}\end{equation}
is a noise built up by the contribution of each white noise with weighted function $\phi_\rho$. Therefore, the variance of the smoothed noise is
\begin{equation*}
\E{(W^H_\phi(\rho))^2}= \sum^{\infty}_{n=1} (\phi_\rho, M_H\xi_n)^2.
\end{equation*}
Since $(\phi_\rho, M_H\xi_n)= c_H\int_\Re d\nu\, \widehat{\phi}(\nu)\abs{\nu}^{1/2-H} \hat{\xi}_n\exp -i \nu\rho,$ it is
\begin{align*}
\sum^{\infty}_{n=1} (\phi_\rho, M_H \xi_n)(\phi_{\rho'}, M_H\xi_n)
&= c^2_H \int_\Re d\nu\, |\widehat{\phi}(\nu)|^2 \abs{\nu}^{1-2H} \exp -i\nu (\rho-\rho')\\ &= \sqrt{2\pi}\,c^2_H \mathscr{F}^{-1}\left[|\widehat{\phi}|^2 \abs{\cdot}^{1-2H}\right](\rho-\rho')
\end{align*}
where $\widehat{\phi}^\ast(\nu)= \widehat{\phi}(-\nu)$ have been used. Finally, the generalized angle-of-arrival variance acquires the form
\begin{equation}
\sigma^2_{m,H}=C_z^2 \sqrt{2\pi} \, c^2_H r_0^{-2H} \mathscr{F}\left[|\widehat{\phi}|^2 \abs{\cdot}^{1-2H}\right](0).
\label{eq:smooth-AOA}\end{equation}
Observe that the function $\phi$ is a distribution-like function, it must satisfy the condition:
\begin{equation}
\int_\Re \phi(s) = \sqrt{2\pi}\, \widehat{\phi}(0)\equiv 1.
\label{eq:norm-cond}\end{equation}  

The natural election for $\widehat{\phi}$ is the Fourier transform of a pupil with diameter $D$. Its normalized version according to Eq.~\eqref{eq:norm-cond} is,   
\begin{equation*}
\widehat{\phi} (\nu)= \sqrt{\frac{2}{\pi}} \frac{J_1(\nu D/2)}{(\nu D/2)}.
\end{equation*}
Therefore, using Eq.~\eqref{eq:smooth-AOA} and the pupil filtering function,
\begin{align}
\sigma^2_{m,H} &= C_z^2 c^2_H r_0^{-2H} \int_\Re\!d\nu\, \abs{\nu}^{1-2H} \frac{2}{\pi}\frac{J^2_1(\nu D/2)}{(\nu D/2)^2}\,\nonumber\\
&= \frac{\Gamma(2H+1)\Gamma(H+1/2)\Gamma(1-H)\sin\pi H}{\pi^{3/2} 2^{2H-3}\Gamma(H+1)\Gamma(H+2)}\,\frac{C^2_\varphi}{2\pi^2}\lambda^2 r_0^{-2H} D^{2H-2}.
\label{eq:pupil}
\end{align}
Finally, it is $\sigma^2_{m,5/6} = 1.04313\, \sigma^2_m$. As we remove high frequencies the noise becomes more regular, and the wave-front variance approaches to that of Tatarsk\u{\i}.

\section{CONCLUSIONS}

This paper introduces a stochastic process, the ifBm, to model the turbulent wave-front phase. Not only it gives the right structure function for non-Kolmogorov turbulence, but also adds well-known statistic properties of the wave-front phase. Moreover, our model allows to extend results for two relevant optical quantities: the Strehl ratio and the angle-of-arrival variance. The expressions for these quantities depend on the Hurst parameter, thus on the dynamic state of the turbulence\cite{paper:bester}. Remember that this parameter is related to the site location where the measurements are made.

In particular, the expression obtained for the angle-of-arrival variance when $H=5/6$, Eq.\eqref{eq:pupil}, is  almost identical to the classical one found by Tatarsk\u{\i} when high frequencies are filtered out. Nonetheless, for a Hurst parameter different from the one above a dependence with the wavelength appears. Up to now, it is unclear for us if such dependence exists for non-Kolmogorov turbulence. That is, if the Fried parameter is independent or not from the Hurst parameter.

Using the formalism presented here a wider range of power spectra can be studied. Such as multifractal processes where the power exponent changes across frequency ranges. 

Also, asymmetric power spectra gives rise to self-affine surfaces; the phase $\varphi$ is scaled differently depending on chosen axis. Thus, two Hurst parameters can control this behavior, and this formalism is applicable again.

Finally, since phase distortions of a wave-front transform into amplitude distortions in the wave cross sections, a similar analysis should be possible for the amplitude.

\section*{APPENDIX A}\label{appx:fbm-calc}
\setcounter{footnote}{0}
\setcounter{equation}{0}
\renewcommand{\theequation}{A.\arabic{equation}}
\setcounter{figure}{0}
\renewcommand{\thefigure}{A.\arabic{figure}}

The purpose of this appendix is not to give a complete exposition of the calculus developed by Elliott and van der Hoek\cite{paper:elliot}, but an introduction of the tools used in this work. 

First, let $M_H$ be an operator defined for any $0<H<1$ such that,
\begin{equation}
\widehat{M_H \phi}(\nu)= c_H \abs{\nu}^{1/2-H} \widehat{\phi}(\nu),
\label{eq:H-operator}\end{equation}
where the hat stands for the Fourier transform, $c_H^2= \Gamma(2H+1)\sin \pi H$, and the function $\phi$ is defined as in Sec.~\ref{sec:fbm}. The generalized fractional white noise is 
\begin{equation}
W^H(\phi)=\langle M_H\phi,\omega\rangle=\int_\Re (M_H\phi)\, dB =\int_\Re \phi\, dB^H;
\label{eq:general-fwn}\end{equation}
thus, the extension to the It\^o integral (with deterministic integrator) is again a bilinear map. Nevertheless, the test function $\phi$ must belong to $L^2_H(\Re)=\{\phi: \widehat{M_H \phi}\in L^2(\Re)\}$. The fractional Brownian motion can be constructed as $B^H(t)(\omega)=\langle M_H \mathbbm{1}_{(0,t)},\omega\rangle$---where $\mathbbm{1}_{(a,b)}(t)$ is the indicator function.\cite{note:index} Two properties of the generalized noise are worth mentioning: $\E{\langle M_H\phi,\omega\rangle}=0$, and the \textit{isometry} 
\begin{equation*}
\E{\langle M_H\phi,\omega\rangle \langle M_H\phi',\omega\rangle} = (M_H\phi,M_H\phi').
\end{equation*}
Afterwards, the stochastic processes subject to the same probability space are defined through what is called \textit{Chaos expansion}. Shortly, any stochastic process $X$ can be written as the formal sums
\begin{equation*}
X(\omega)=\sum_\alpha c_\alpha\, \mathcal{H}_\alpha(\omega),
\end{equation*}
with $c_\alpha^2 =\E{X \mathcal{H}_\alpha}/\alpha!$. Here it is defined $\alpha!=\alpha_1!\alpha_2!\dots\alpha_n!$, the factorial of the finite non-negative integer multi-index $\alpha$. While  $\mathcal{H}_\alpha(\omega)=\prod^n_{i=1} H_{\alpha_i}(\langle\xi_i,\omega\rangle)$ represents the stochastic component of the process, and it is build up through the \textit{Hermite functions}: 
\begin{equation}
\xi_n(x) = \frac{e^{-x^2/2}H_{n-1}(x)}{\sqrt{2^{n-1}(n-1)! \pi^{1/2}}},
\end{equation}
with $H_n$ the \textit{Hermite polynomials}. These functions form an orthogonal basis satisfying:
\begin{equation*}
\sum^\infty_{n=1}\xi_n(x)\xi_n(y)=\delta(x-y),\quad\text{ and }\quad\int_\Re\!\!dx\, \xi_n(x)\xi_m(x)=\delta_{n,m}.
\end{equation*}
Particularly, consider that $M_H\mathbbm{1}_{[0,t]}(s)=\sum^\infty_{n=1}(M_H\mathbbm{1}_{[0,t]},\xi_n)\,\xi_n(s)$, then
\begin{equation}
B^H(t)(\omega)=\langle M_H\mathbbm{1}_{[0,t]},\omega\rangle=\sum^\infty_{n=1}\;(M_H\mathbbm{1}_{[0,t]},\xi_n)\,\langle\xi_n,\omega\rangle.
\end{equation}
This is the chaos expansion for the fBm, since it is univocally defined and the  $\langle\xi_n,\omega\rangle=\mathcal{H}_{\epsilon_n}(\langle\xi_n,\omega\rangle)$, please refer to Elliott and van der Hoek. Moreover, the operator $M_H$ is self-adjoint, so 
\begin{equation*}
(M_H\mathbbm{1}_{[0,t]},\xi_n)=(\mathbbm{1}_{[0,t]},M_H\xi_n) =\int^t_0\! ds\, M_H\xi_n(s); 
\end{equation*}
the fractional white noise has the expansion
\begin{equation}
\frac{d}{dt}B^H(t)=\sum^\infty_{n=1} M_H\xi_n(t)\langle\xi_n,\omega\rangle= W^H(t).
\label{eq:fwn-exp}\end{equation}
The stochastic integral in equation \eqref{eq:general-fwn} is then rewritten as
\begin{align}
W^H(\phi) =& \sum^\infty_{n=1} (\phi, M_H\xi_n) \langle\xi_n,\omega\rangle\nonumber\\
=& \int_\Re\! ds\, \phi(s) \left[\sum^\infty_{n=1} M_H\xi_n(s) \langle\xi_n,\omega\rangle\right]\nonumber\\
=& \int_\Re \phi(s)\, W^H(s)\, ds.
\end{align}
The fractional Brownian noise is the integrated version of the generalized one. Also, the \textit{smoothed} noise is defined: $W^H_\phi(t)=  \int_\Re \phi(s-t)\, W^H(s)\, ds.$ 

Finally, it remains to be introduced the Wick product. Suppose $X(\omega)=\sum_\alpha a_\alpha\mathcal{H}_\alpha(\omega)$ and $Y(\omega)=\sum_\beta b_\beta\mathcal{H}_\beta(\omega)$ then the product is
\begin{equation*}
(X \diamond Y )(\omega)= \sum_{\alpha,\beta} a_\alpha b_\beta \mathcal{H}_{\alpha+\beta}(\omega).
\end{equation*}
Among the properties it has, maybe the most relevant is $\E{X\diamond Y}=\E{X}\cdot\E{Y}$. For the particular case $X(\omega)=a_0 + \sum_{n=1}^\infty a_n \mathcal{H}_{\epsilon_n}(\omega)$ and $Y(\omega)= b_0 + \sum_{m=1}^\infty a_m \mathcal{H}_{\epsilon_m}(\omega)$ is
\begin{equation}
(X\diamond Y)(\omega) = X(\omega) Y(\omega) -\sum_{n=1}^\infty a_n b_n.
\label{eq:wick-gauss}\end{equation}
\section*{APPENDIX B}\label{appxA}
\setcounter{footnote}{0}
\setcounter{equation}{0}
\renewcommand{\theequation}{B.\arabic{equation}}
\setcounter{figure}{0}
\renewcommand{\thefigure}{B.\arabic{figure}}

The variance of the increments of an isotropic fractional Brownian motion is, using Eq.~\eqref{eq:fbm-cov}:
\begin{equation}
\E{\left(\tilde{B}^H(\brho) -\tilde{B}^H(\brho') \right)^2} =\abs{\rho- \rho'}^{2H}.
\label{eq:cov-ex}\end{equation}
It will be shown that when $\brho'$ and $\brho$ are `near' enough the latter turns into
\begin{equation}
\E{\left(\tilde{B}^H(\brho) -\tilde{B}^H(\brho') \right)^2} \simeq \norm{\brho- \brho'}^{2H}.
\label{eq:cov-app}\end{equation}
It can be seen from Fig.~\ref{fig:near-vects} that $\brho - \brho'= \Bs{\delta}_\theta + \Bs{\delta}_r$ and $ \abs{\rho- \rho'}^2=\delta^2=\norm{\Bs{\delta}_r}^2$. Thus, Eqs.~\eqref{eq:cov-ex} and \eqref{eq:cov-app} can be compared as follows
\begin{align*}
\frac{\norm{\brho-\brho'}^{2H} - \norm{\Bs{\delta}_r}^{2H}}{\delta^{2H}} &= \frac{\norm{ \Bs{\delta}_\theta + \Bs{\delta}_r}^{2H} - \delta^{2H}}{\delta^{2H}} \\
&= \left(1+ \frac{\delta_\theta^2 + 2 \Bs{\delta}_\theta \cdot\Bs{\delta}_r }{\delta^2}\right)^H -1\\
&= H \left( \frac{\delta_\theta^2 + 2 \Bs{\delta}_\theta \cdot\Bs{\delta}_r }{\delta^2}\right) + \mathcal{O}\left( \frac{\delta_\theta^2 + 2 \Bs{\delta}_\theta \cdot\Bs{\delta}_r }{\delta^2}\right)^2.
\end{align*}
Therefore, the condition
\begin{equation*}
\frac{\delta_\theta^2 + 2 \Bs{\delta}_\theta \cdot\Bs{\delta}_r}{\delta^2}\ll 1
\end{equation*}
must be fullfiled to allow the replacement made at Eq.~\eqref{eq:cov-app}. Setting $\brho'= \rho (\cos\theta,\sin\theta)$ and $\brho=(\rho,0)$, it is $\Bs{\delta}^2_\theta= 2 \rho^2 (1-\cos\theta)$ and $2 \Bs{\delta}_\theta \cdot\Bs{\delta}_r = 2\delta\rho(1-\cos\theta)$. Supposing $\theta\ll 1$ and $\delta\ll 1$, 
\begin{equation*}
\frac{\delta_\theta^2 + 2 \Bs{\delta}_\theta \cdot\Bs{\delta}_r}{\delta^2} \sim \theta^2 \left(\frac{\rho^2}{\delta^2} + \frac{\rho}{\delta}\right) \sim \theta^2\rho^2\delta^{-2},
\end{equation*}
If it is set $\delta\sim \delta^{-2}\delta^2_\theta$, then the condition
\begin{equation}
\delta< \delta\,\delta^{1/2} \sim \delta_\theta \ll 1
\label{eq:less-1/2}\end{equation}
arises verifing Eq.~\eqref{eq:cov-app}.

\section*{ACKNOWLEDGMENTS}

DGP has been supported financially by a postdoctoral research fellowship from the Innovatec Foundation (Argentina) and LZ by a research fellowship from the Consejo Nacional de Investigaciones Cient\'{\i}ficas y T\'ecnicas (CONICET, Argentina).

Dar\'{\i}o G. P\'erez, Luciano Zunino and Mario Garavaglia can be reached by e-mail at \textsf{dariop@ciop.unlp.edu.ar}, \textsf{lucianoz@ciop.unlp.edu.ar} and \textsf{garavagliam@ciop.unlp.edu.ar}, respectively.

\bibliography{references,footnotes}

\begin{thebibliography}{10}
\bibitem{note:frequency}
{In this paper instead of the usual frequency $f$, $\nu=2\pi f$} is used for
  self-consistence.
\bibitem{note:tilt}
{If tilt aberration is present, the axis of this definition would be normal to
  the plane of that tilt.}
\bibitem{note:ft}
{In this paper the Fourier transform is defined as
  $\widehat{\phi}(\nu)=\tfrac{1}{\sqrt{2\pi}}\int_\Re\!ds\, \phi(s)\exp -i\nu
  s$.}
\bibitem{note:index}
{It is equal to $1$ if $t$ falls in the interval $(a,b)$, zero otherwise, and
  $-1$ when $b<a$}.
\end{thebibliography}
\bibliographystyle{unsrt}

\newpage

\begin{figure}
\psset{unit=1cm}
\begin{center}
\begin{pspicture}(0,0)(4,5.5)
\psline[linewidth=1pt,arrowsize= 2.5pt 2.5]{->}(0,0)(2,3.46)
\psline[linewidth=1pt,arrowsize= 2.5pt 2.5]{->}(0,0)(3.46,2)
\psline[linewidth=1pt,arrowsize= 2.5pt 2.5]{->}(2,3.46)(2.75,4.76)
\psline[linewidth=1pt,arrowsize= 2.5pt 2.5]{->}(3.46,2)(2,3.46)
\psline[linewidth=1pt,arrowsize= 2.5pt 2.5]{->}(3.46,2)(2.75,4.76)
\psarc[arcsep=1pt,linewidth=.5pt]{<-}(0,0){1.5}{30}{60}
\psarc[linewidth=.5pt,linestyle=dotted]{-}(0,0){4}{25}{65}
\psbrace[linewidth=.5pt,nodesepA=-10pt,nodesepB=5pt,ref=lt,braceWidth=.5,rot=-150](2.75,4.76)(0,0){$\brho$}
\uput[ur](1,1){$\theta$}
\uput[dr](1.3,2.5){$\brho'_\theta$}
\uput[dr](1.73,1){$\brho'$}
\uput[dr](2.1,3){$\Bs{\delta}_\theta$}
\uput[dr](2.1,4){$\Bs{\delta}_r$}
\uput[dr](3.3,3.7){$\Delta\brho = \brho -\brho'$}
\psdots[dotstyle=+,dotscale=2,border=1pt](1.51,3.49)
\psdots[dotstyle=|,dotscale=2,dotangle=45,border=1pt](1.5,3.56)
\end{pspicture}
\end{center}
\caption{Decomposition of the vector $\brho - \brho'$ in terms of $\Bs{\delta}_\theta$ and $\Bs{\delta}_r$.}
\label{fig:near-vects}
\end{figure}
\end{document}